\documentclass[english,reprint,a4paper,aps,pre]{revtex4-1}
\usepackage[T1]{fontenc}
\usepackage[latin9]{inputenc}
\usepackage{amsbsy}
\usepackage{amstext}
\usepackage{graphicx}
\usepackage{esint}
\usepackage{babel}
\begin{document}

\title{Maximum entropy principle for stationary states underpinned by stochastic
thermodynamics}

\author{Ian J. Ford }

\affiliation{Department of Physics and Astronomy and London Centre for Nanotechnology,
University College London, Gower Street, London WC1E 6BT, U.K.}
\begin{abstract}
The selection of an equilibrium state by maximising the entropy of
a system, subject to certain constraints, is often powerfully motivated
as an exercise in logical inference, a procedure where conclusions
are reached on the basis of incomplete information. But such a framework
can be more compelling if it is underpinned by dynamical arguments,
and we show how this can be provided by stochastic thermodynamics,
where an explicit link is made between the production of entropy and
the stochastic dynamics of a system coupled to an environment. The
separation of entropy production into three components allows us to
select a stationary state by maximising the change, averaged over
all realisations of the motion, in the principal relaxational or nonadiabatic
component, equivalent to requiring that this contribution to the entropy
production should become time independent for all realisations. We
show that this recovers the usual equilibrium probability density
function (pdf) for a conservative system in an isothermal environment,
as well as the stationary nonequilibrium pdf for a particle confined
to a potential under nonisothermal conditions, and a particle subject
to a constant nonconservative force under isothermal conditions. The
two remaining components of entropy production account for a recently
discussed thermodynamic anomaly between over- and underdamped treatments
of the dynamics in the nonisothermal stationary state.
\end{abstract}
\maketitle

\section{Introduction}

The standard distributions in equilibrium statistical mechanics can
be derived in an appealingly straightforward fashion using the principle
of maximum entropy or MaxEnt. The procedure seems first to have been
employed by Gibbs \cite{Gibbs-book1902}, and was vigorously championed
by Jaynes \cite{Jaynes03} as an example of logical inference, namely
the optimal determination of a statistical description of an imperfectly
specified system.

The essential idea is that a system can possess an informational entropy
that measures the uncertainty of an observer's perception, expressed
through a probability distribution over all the configurations (microstates)
that the system could adopt. In order to represent the situation in
as objectively neutral a fashion as possible, so the argument goes,
we should select the distribution that has the greatest informational
entropy, while enforcing consistency with any known properties of
the system, represented as expectation values over the distribution.
For various mathematical and physical reasons \cite{Khinchin-book,Grandy-book2012},
the informational entropy $S_{I}$ of a probability distribution $p(i)$
over a set of microstates labelled $i$ is written $S_{I}=-\sum_{i}p(i)\ln p(i)$
and if the constraints take the form of fixed expectation values $\langle C_{n}\rangle=\sum_{i}C_{n}(i)p(i)$
of a set of microstate-dependent quantities $C_{n}$, then it can
be shown by way of the method of Lagrange multipliers that the statistical
representation that makes no unwarranted further assumptions about
the system is $p(i)\propto\exp[-\sum_{n}\lambda_{n}C_{n}(i)]$ where
the $\lambda_{n}$ are constants. So if we consider a physical system
and employ the constraint that it possesses an identifiable mean energy
as a consequence of being coupled to an isothermal reservoir acting
as a source and sink of heat, then the optimal representation is canonical:
$p(i)\propto\exp(-E(i)/kT_{r})$, where $E(i)$ is the system energy
in microstate $i$, the reservoir is characterised by a temperature
$T_{r}$, and $k$ is the Boltzmann constant. Such a distribution
would clearly represent a situation with time independent system properties.

Jaynes argued that similar procedures should be used to select probability
density functions (pdfs) in more general situations, particularly
for nonequilibrium stationary states \cite{Jaynes79,Jaynes03}. The
general strategy would be to introduce constraints relating to the
existence of a non-zero flux of energy or particles through or within
a system and some progress along these lines has been made, for example
in \cite{EvansRML04}. Principles for selecting the most probable
path taken by a system have a long history (\cite{Filyokov67,Monthus11}
also reviewed in \cite{Martyushev06}) and a similar rationale underlies
the nonequilibrium statistical operator method (NESOM) developed by
Zubarev and coworkers \cite{Zubarev74,Luzzi00,Luzzi02}. Variational
principles involving the production \emph{rate} of entropy in stationary
states have also been proposed \cite{Prigogine-book1961,Dewar14}
as well as the maximisation of \emph{relative} entropy \cite{Banavar10}.

A puzzling aspect of MaxEnt, however, is that the selection of the
prevailing constraints appears to be rather arbitrary. For example,
if a system is exposed to a heat reservoir such that an expectation
value of energy is identifiable and therefore relevant to the MaxEnt
procedure, then why are the expectation values of additional functions
of energy not relevant? Why does a term proportional to $[E(i)]^{2}$
not appear in the exponent of the canonical distribution in addition
to $E(i)$? An argument is often made that constraints are placed
upon dynamically conserved quantities, which would exclude arbitrary
functions of energy, but this strongly suggests that dynamics must
underpin the procedure. It is our aim here to demonstrate that the
framework of stochastic thermodynamics can provide the underlying
dynamics in the derivation of the MaxEnt principle.

In Section \ref{sec:Stochastic-dynamics-and} we give a brief overview
of stochastic thermodynamics, discussing the way in which (stochastic)
entropy production is both a reflection of the mechanical irreversibility
of the underlying stochastic dynamics and the basis of a measure of
the change in microstate-level uncertainty of a system with time.
We describe how the total entropy production may usefully be separated
into three components, each with a specific character. In Section
\ref{sec:Dynamical-MaxEnt-selection} we discuss the dynamics of mean
total entropy production for a system subject to a conservative force
field and coupled to an isothermal reservoir, arguing that this viewpoint
provides a route to the canonical solution to the stochastic dynamics,
and showing how this maps onto the traditional MaxEnt variational
principle. We see how the constraints on the variational procedure
emerge as a result of the dynamical coupling of the system to the
reservoir. Only one of the three components of entropy production
is non-zero in this case.

In Section \ref{sec:nonisothermal} we consider a nonequilibrium situation
where a system is exposed to a background temperature gradient and
show that the stationary state may be selected by the maximisation
of the mean relaxational or nonadiabatic entropy production, one of
the three components, or equivalently by requiring the increment in
this component to be zero for all possible dynamical scenarios. We
note that the average of the remaining two components accounts for
a recently discussed anomaly in entropy production between over- and
underdamped treatments \cite{Celani12,Ge14}. Supporting analysis
is provided in Appendix \ref{sec:Components-of-stochastic} and a
system driven by a nonconservative force under isothermal conditions
is discussed in Appendix \ref{sec:Particle-in-a}. We give our conclusions
in Section \ref{sec:Conclusions}.

\section{Stochastic dynamics and thermodynamics\label{sec:Stochastic-dynamics-and}}

Stochastic thermodynamics is based on a set of stochastic differential
equations (SDEs) that describe the evolution of system dynamical variables
\cite{Risken89,vanKampen07,Gardiner09}, together with a definition
of the entropy production associated with a possible realisation of
the motion \cite{sekimoto1,sekimoto2,seifertoriginal,seifertprinciples}.

We focus our discussion on the motion in one spatial dimension of
a single particle coupled to a set of heat reservoirs, each corresponding
to a given position of the particle. The motion is described by the
following It$\bar{{\rm o}}$-rules SDEs:
\begin{eqnarray}
dx & = & vdt\label{eq:1}\\
dv & = & -\gamma vdt+\frac{F(x)}{m}dt+\left(\frac{2kT_{r}(x)\gamma}{m}\right)^{1/2}dW,\label{eq:1a}
\end{eqnarray}
where $x$ and $v$ are the particle position and velocity, respectively,
$t$ is time, $\gamma$ is the friction coefficient, $F(x)$ is a
spatially dependent force field acting on the particle, assumed for
the moment to be related to a potential $\phi(x)$, $m$ is the particle
mass, $T_{r}(x)$ is a spatially dependent reservoir temperature and
$dW$ is an increment in a Wiener process. Such a starting point for
discussing the stochastic behaviour of a particle in a temperature
gradient is often employed, though alternatives can also be imagined.
The simplicity of Eqs. (\ref{eq:1}) and (\ref{eq:1a}) is convenient
for our purpose.

Given the dynamics, entropy production is defined in the fashion proposed
by Seifert \cite{seifertoriginal}. It is fundamentally a measure
of the probabilistic mechanical irreversibility of the motion. For
a given time interval $0\le t\le\tau$, the dynamics can generate
a trajectory $\vec{\boldsymbol{x}},\vec{\boldsymbol{v}}$ (where $\vec{\boldsymbol{x}}$
represents a function $x(t)$ in the specified time interval) according
to a probability density function ${\cal P}[\vec{\boldsymbol{x}},\vec{\boldsymbol{v}}]$.
In the situation under consideration, the latter can be written as
a product of the probability density $p(x,v,t)$ of a microstate with
$x=x(0)$ and $v=v(0)$ at $t=0$, and a conditional probability density
that the specified trajectory is followed thereafter. The dynamics
are also capable of generating an \emph{antitrajectory} after an inversion
of the particle velocity at time $\tau$, and under the influence
of a reversed time evolution of the force field and reservoir temperature,
if relevant \cite{Ford-book2013,SpinneyFordChapter13,Ford15}, until
a total time $2\tau$ has elapsed. In this period $\tau\le t\le2\tau$
we can identify the probability density ${\cal P}^{{\rm R}}[\vec{\boldsymbol{x}}^{\dagger},\vec{\boldsymbol{v}}^{\dagger}]$
that an antitrajectory $\vec{\boldsymbol{x}}^{\dagger},\vec{\boldsymbol{v}}^{\dagger}$
starting at $x(\tau),-v(\tau)$ and ending at $x(0),-v(0)$ is generated,
with the superscript R reminding us that the potential and reservoir
temperature evolve with time in a reverse fashion with respect to
the period $0\le t\le\tau$. The antitrajectory $\vec{\boldsymbol{x}}^{\dagger},\vec{\boldsymbol{v}}^{\dagger}$
is the `time-reversed' partner of $\vec{\boldsymbol{x}},\vec{\boldsymbol{v}}$
\cite{SpinneyFord12a,SpinneyFord12b,SpinneyFord12c}. The total entropy
production associated with the trajectory $\vec{\boldsymbol{x}},\vec{\boldsymbol{v}}$
is then defined by
\begin{equation}
\Delta s_{{\rm tot}}[\vec{\boldsymbol{x}},\vec{\boldsymbol{v}}]=\ln\left[\frac{{\cal P}[\vec{\boldsymbol{x}},\vec{\boldsymbol{v}}]}{{\cal P}^{{\rm R}}[\vec{\boldsymbol{x}}^{\dagger},\vec{\boldsymbol{v}}^{\dagger}]}\right],\label{eq:2}
\end{equation}
and after multiplication by Boltzmann's constant and averaging over
all trajectories, this corresponds to the production of thermodynamic
entropy in the process. In a condition of thermal equilibrium, when
the dynamics would be expected to generate a trajectory and its time-reversed
partner with equal likelihood, the entropy production associated with
\emph{all} feasible trajectories will vanish.

The entropy production along a trajectory evolves stochastically,
and for the system dynamics considered here an increment in $\Delta s_{{\rm tot}}$
is specified by the It$\bar{{\rm o}}$-rules SDE
\begin{equation}
d\Delta s_{{\rm tot}}=-d[\ln p(x,v,t)]-\frac{1}{kT_{r}(x)}d\left(\frac{mv^{2}}{2}\right)+\frac{F(x)}{kT_{r}(x)}dx.\label{eq:3}
\end{equation}
The origin of this expression is described in Appendix \ref{sec:Components-of-stochastic}
and elsewhere \cite{seifertoriginal,SpinneyFord12b}. The second term
is the negative increment in the kinetic energy of the particle over
the time interval $dt$, and the third term is the negative increment
in its potential energy, both divided by the local reservoir temperature.
Together, they represent a positive increment in the energy of the
local reservoir (which we may regard as a heat transfer $dQ_{r}$
to that reservoir) divided by the local temperature. This would then
correspond to a Clausius-type incremental change $d\Delta s_{{\rm res}}=dQ_{r}/kT_{r}(x)$
in the entropy of the local reservoir in the interval of time $dt$.
 It is then natural to regard the first term in Eq. (\ref{eq:3})
as the change in the entropy of the system (the particle) over this
period. Seifert defined a stochastic system entropy $s_{{\rm sys}}=-\ln p(x,v,t)$
in terms of the evolving phase space probability density function
$p$ generated by the stochastic dynamics \cite{seifertoriginal},
such that we can write
\begin{equation}
d\Delta s_{{\rm tot}}=d\Delta s_{{\rm sys}}+d\Delta s_{{\rm res}}.\label{eq:4}
\end{equation}

Since velocity evolves in Eq. (\ref{eq:1a}) under the direct influence
of a stochastic force, the rules of stochastic calculus apply when
we manipulate increments of a function of $v$. Since we employ It$\bar{{\rm o}}$
rules it would be incorrect to proceed from Eq. (\ref{eq:3}) by writing
$d(v^{2})=2vdv,$ which only applies under Stratonovich rules. An
additional term proportional to $dt$ would appear \cite{Gardiner09,Sokolov10}.
Taking this properly into account removes an apparent anomaly in $d\Delta s_{{\rm tot}}$
discussed in \cite{Luposchainsky13}.

The evaluation of $\Delta s_{{\rm tot}}$ for a specific realisation
of the motion requires us to determine the evolution of the pdf as
well as the system variables $x$ and $v$: we need to solve the Fokker-Planck
equation \cite{Risken89}
\begin{equation}
\frac{\partial p}{\partial t}={\cal L}p=-\frac{\partial J_{v}^{{\rm ir}}}{\partial v}-v\frac{\partial p}{\partial x}-\frac{F}{m}\frac{\partial p}{\partial v},\label{eq:fpe}
\end{equation}
that corresponds to Eqs. (\ref{eq:1}) and (\ref{eq:1a}), where $J_{v}^{{\rm ir}}=-\gamma vp-\partial(D_{v}p)/\partial v$
is the irreversible probability current, with $D_{v}=\gamma kT_{r}(x)/m$.

In spite of fluctuations in the total entropy production as the particle
follows a trajectory, it can be shown that the average of this quantity
is non-negative, a property that arises from an integral fluctuation
relation \cite{seifertoriginal}. This is regarded as the second law
of thermodynamics in this framework, expressed as $d\langle\Delta s_{{\rm tot}}\rangle=d\langle\Delta s_{{\rm sys}}\rangle+d\langle\Delta s_{{\rm res}}\rangle\ge0$
 where the angled brackets denote an expectation over the pdfs of
system coordinates at the beginning and end of the incremental time
period.

The entropy production can be separated into a specific set of components,
each with a particular character. Initial developments in this direction
were provided by Van den Broeck and Esposito \cite{adiabaticnonadiabatic0,adiabaticnonadiabatic1,adiabaticnonadiabatic2}
and extended by Spinney and Ford \cite{SpinneyFord12a,SpinneyFord12b,SpinneyFord12c},
working within a framework suggested by Oono and Paniconi \cite{oono}.
The total entropy production may be written as three terms \cite{SpinneyFord12a,SpinneyFord12b}
\begin{equation}
d\Delta s_{{\rm tot}}=d\Delta s_{1}+d\Delta s_{2}+d\Delta s_{3},\label{eq:5a}
\end{equation}
with the $\Delta s_{1}$ and $\Delta s_{2}$ components defined in
terms of ratios of probabilities that specific trajectories are taken
by the system. We give particular attention to the first component,
given by
\begin{equation}
\Delta s_{1}[\vec{\boldsymbol{x}},\vec{\boldsymbol{v}}]=\ln\left[\frac{{\cal P}[\vec{\boldsymbol{x}},\vec{\boldsymbol{v}}]}{{\cal P}^{{\rm ad,R}}[\vec{\boldsymbol{x}}^{R},\vec{\boldsymbol{v}}^{R}]}\right],\label{eq:5b}
\end{equation}
where $\vec{\boldsymbol{x}}^{R},\vec{\boldsymbol{v}}^{R}$ represents
a reversal of the system trajectory \emph{without} the inversion of
velocity coordinates \cite{SpinneyFord12b}, and the superscript `ad'
indicates that `adjoint' dynamical rules are employed to work out
the probability of its generation.

Our key point is that a dynamical underpinning of the MaxEnt principle
can be obtained by considering the properties of the $\Delta s_{1}$
component. The approach of the pdf under the dynamics towards stationarity
is equivalent to a variational principle for its selection expressed
in terms of $\Delta s_{1}$. Such a principle holds irrespective of
whether the stationary state is one of equilibrium, in which case
$\Delta s_{2}$ and $\Delta s_{3}$ are both zero during the evolution,
or nonequilibrium such that $\Delta s_{2}$ and $\Delta s_{3}$ are
in general non-zero. In both cases $d\Delta s_{1}$ vanishes asymptotically
and $\langle\Delta s_{1}\rangle$ reaches a ceiling.

Let us substantiate these claims. It was shown in \cite{SpinneyFord12a,SpinneyFord12b}
that the average values of $\Delta s_{1-3}$ are related to the transient
and stationary system pdfs ($p$ and $p_{{\rm st}}$, respectively)
as follows:
\begin{eqnarray}
\frac{d\langle\Delta s_{1}\rangle}{dt} & = & -\int dxdv\;\frac{\partial p}{\partial t}\ln\left[\frac{p(x,v,t)}{p_{{\rm st}}(x,v)}\right]\nonumber \\
 & = & \int dxdv\;\frac{p}{D_{v}}\left(\frac{J_{v}^{{\rm ir}}}{p}-\frac{J_{v}^{{\rm ir,st}}}{p_{{\rm st}}}\right)^{2}\ge0,\label{S1av2}\\
\frac{d\langle\Delta s_{2}\rangle}{dt} & = & \int dxdv\;\frac{p}{D_{v}}\left[\frac{J_{v}^{{\rm ir,st}}(x,-v)}{p_{{\rm st}}(x,-v)}\right]^{2}\ge0,\label{eq:s2av}\\
\frac{d\langle\Delta s_{3}\rangle}{dt} & = & -\int dxdv\;\frac{\partial p}{\partial t}\ln\left[\frac{p_{{\rm st}}(x,v)}{p_{{\rm st}}(x,-v)}\right],\label{eq:s3av}
\end{eqnarray}
where ${\cal L}p_{{\rm st}}=0$, and $J_{v}^{{\rm ir,st}}=-\gamma vp_{{\rm st}}-\partial(D_{v}p_{{\rm st}})/\partial v$
is the irreversible probability current in the stationary state. The
positivity of the rate of change of $\langle\Delta s_{1}\rangle$
and $\langle\Delta s_{2}\rangle$ is here explicit, but is also a
consequence of integral fluctuation relations for these two components
of entropy production \cite{hatanosasa,adiabaticnonadiabatic0,SpinneyFord12a}.
Furthermore, the unaveraged increments $d\Delta s_{1}$ and $d\Delta s_{3}$
take the form
\begin{equation}
d\Delta s_{1}=-d[\ln p(x,v,t)]+d[\ln p_{{\rm st}}(x,v)],\label{eq:ds1}
\end{equation}
and
\begin{equation}
d\Delta s_{3}=-d[\ln p_{{\rm st}}(x,v)]+d[\ln p_{{\rm st}}(x,-v)],\label{eq:ds3}
\end{equation}
making clear the conditions for which they vanish ($p(x,v,t)=p_{{\rm st}}(x,v)$
and $p_{{\rm st}}(x,v)]=p_{{\rm st}}(x,-v)$, respectively), while
$d\Delta s_{2}=d\Delta s_{{\rm res}}-d[\ln p_{{\rm st}}(x,-v)]$.
The SDEs that govern the evolution of $\Delta s_{1-3}$ for a general
Markovian dynamical framework are given in Appendix \ref{sec:Components-of-stochastic}.

The three contributions to the total entropy production can be interpreted
as follows. $\Delta s_{1}$ is the principal relaxational entropy
production associated with the approach of a system towards a stationary
state. Once a system is in a stationary state, with $p=p_{{\rm st}}$,
no further increments in $\Delta s_{1}$ take place. This component
of entropy production, averaged over all possible realisations of
the motion initiated at $t=0$, namely $\langle\Delta s_{1}\rangle$,
increases monotonically towards a positive constant, since Eq. (\ref{S1av2})
indicates that $d\langle\Delta s_{1}\rangle/dt\to0$ as $p\to p_{{\rm st}}$.
The $\Delta s_{1}$ contribution was denoted the nonadiabatic entropy
production by Esposito and Van den Broeck \cite{adiabaticnonadiabatic0,adiabaticnonadiabatic1,adiabaticnonadiabatic2}
and its properties were given particular attention by Hatano and Sasa
\cite{hatanosasa}.

$\Delta s_{3}$ is also associated with relaxation towards the stationary
state, but in contrast to $\Delta s_{1}$, its average value does
not necessarily evolve monotonically with time; there is no definite
sign attached to $d\langle\Delta s_{3}\rangle/dt$ in Eq. (\ref{eq:s3av}).
If the stationary pdf is velocity symmetric, however, Eq. (\ref{eq:ds3})
shows that $d\Delta s_{3}$ is identically zero. Since a velocity
asymmetric stationary pdf is typically associated with a non-zero
mean flux of some kind, and also with an underlying breakage of the
principle of detailed balance in the stochastic dynamics \cite{adiabaticnonadiabatic0,adiabaticnonadiabatic1,adiabaticnonadiabatic2},
this component arises in situations where there is a nonequilibrium
stationary state. It was designated the transient housekeeping entropy
production by Spinney and Ford \cite{SpinneyFord12a}.

$\Delta s_{2}$ is also associated with a nonequilibrium stationary
state, since a non-zero rate of change of its average over all possible
realisations of the dynamics requires there to be a non-zero current
$J_{v}^{{\rm ir,st}}$ in the stationary state. The rate of change
of the average $\Delta s_{2}$ is non-negative \cite{IFThousekeeping},
but is non-zero in a nonequilibrium stationary state, in contrast
to the rate of change of the average $\Delta s_{1}$ which would then
be zero. This is the most distinctive difference between these two
components of entropy production. Esposito and Van den Broeck referred
to $\Delta s_{2}$ as the adiabatic entropy production (and they considered
it only in the context of the dynamics of even coordinates such as
position), and Spinney and Ford, who considered odd coordinates such
as velocity as well, denoted it the generalised housekeeping entropy
production.

Other separations of total entropy production into three components
are possible \cite{Lee13}, but the choice employed here has the advantage
that the $\Delta s_{3}$ component vanishes in the absence of velocity
variables in the dynamics, and on average its rate of change in a
stationary state is also zero.

As a system approaches stationarity, all three kinds of entropy production
take place, but in the stationary state only $\Delta s_{2}$ and $\Delta s_{3}$
can potentially receive increments. The mean total entropy production
rate in the stationary state is represented by $d\langle\Delta s_{2}\rangle/dt$
alone. If detailed balance holds, both $\Delta s_{2}$ and $\Delta s_{3}$
would be zero. Our proposal is that the monotonic increase in $\langle\Delta s_{1}\rangle$
towards a ceiling, or equivalently the vanishing of $d\Delta s_{1}$
in the stationary state, underpins MaxEnt for equilibrium situations,
and offers an extension of the procedure to nonequilibrium circumstances.
We now explore the implications of this viewpoint for isothermal and
then nonisothermal situations.

\section{Selection of an equilibrium state\label{sec:Dynamical-MaxEnt-selection}}

If the reservoirs were isothermal ($T_{r}(x)=T_{0}$), then irrespective
of the initial conditions, the system under consideration should relax
under the stochastic dynamics to an equilibrium state with zero irreversible
probability current and canonical statistics. Let us examine how this
works out dynamically and thermodynamically.

We presume that the contributions $d\Delta s_{2}$ and $d\Delta s_{3}$
are zero throughout as a consequence of the associated condition of
detailed balance in the dynamics, as suggested by the velocity symmetry
of the expected canonical pdf. This can be checked later. If the change
in total entropy production is indeed entirely given by $d\Delta s_{1}$
then the total entropy production satisfies not only the incremental
representation in terms of system and reservoir contributions in Eqs.
(\ref{eq:3}) and (\ref{eq:4}), but also the mean behaviour represented
by Eq. (\ref{S1av2}).

We are therefore in a position to write
\begin{equation}
\frac{d\langle\Delta s_{{\rm tot}}\rangle}{dt}\ge0\quad{\rm and}\quad\lim_{t\to\infty}\frac{d\langle\Delta s_{{\rm tot}}\rangle}{dt}=0.\label{eq:6}
\end{equation}
The mean total change in entropy of system plus reservoir will increase
monotonically until it reaches a ceiling. The equilibrium state will
be achieved when $\langle\Delta s_{{\rm tot}}\rangle$ becomes time-independent.

We argue that Eqs. (\ref{eq:6}) are equivalent to the maximisation
of a constrained Gibbs system entropy over the range of possible pdfs,
namely a MaxEnt procedure. The maximisation would have its origin
in the dynamics, and would not arise merely from considerations of
logical inference.

To support this viewpoint, we note that the incremental transfer of
energy $dQ_{r}$ to the reservoir, that specifies the entropy production
$d\Delta s_{{\rm res}}$ in Eq. (\ref{eq:4}), is equal to $-d\Delta E$,
the negative of the increment in the change (with respect to $t=0$)
in the energy of the system as it evolves over the time interval $dt$.
Since $d\Delta s_{{\rm res}}=dQ_{r}/kT_{0}$, we can rewrite Eq. (\ref{eq:4})
as
\begin{equation}
\frac{d\langle\Delta s_{{\rm tot}}\rangle}{dt}=\frac{d\langle\Delta s_{{\rm sys}}\rangle}{dt}-\frac{1}{kT_{0}}\frac{d\langle\Delta E\rangle}{dt}.\label{eq:7}
\end{equation}
Next, we recognise that since $s_{{\rm sys}}=-\ln p$ is a function
of system coordinates at a specified time $t$ rather than a function
of a trajectory of coordinates, the average $\langle\Delta s_{{\rm sys}}\rangle$
over realisations of the dynamics is a difference in the expectation
of $s_{{\rm sys}}$ between the final and initial states. That is
$\langle\Delta s_{{\rm sys}}\rangle=\overline{s_{{\rm sys}}}(t)-\overline{s_{{\rm sys}}}(0)$
where the expectations (indicated by overbars) are averages over the
system pdf; thus $\overline{s_{{\rm sys}}}(t)=\int dxdv\,p(x,v,t)[-\ln p(x,v,t)]=S_{G}(t)$,
the Gibbs informational entropy. For similar reasons, $\langle\Delta E\rangle=\overline{E}(t)-\overline{E}(0)$.
Equations (\ref{eq:6}) and (\ref{eq:7}) tell us that the quantity
$\overline{s_{{\rm sys}}}(t)-\overline{E}(t)/kT_{0}$ increases with
time and reaches a ceiling as a consequence of the exploration of
system phase space represented by the stochastic dynamics and the
evolution of the pdf of the system.

Since Eq. (\ref{eq:7}) may be written $d\langle\Delta s_{{\rm tot}}\rangle/dt=dS_{G}(t)/dt-(kT_{0})^{-1}d\overline{E}(t)/dt$,
the dynamical increase and saturation of $\langle\Delta s_{{\rm tot}}\rangle$,
irrespective of the initial pdf, is, we argue, equivalent to the functional
maximisation of the quantity $S_{G}-\overline{E}/kT_{0}$ with respect
to the pdf. In other words, the equilibrium pdf can also result from
implementing the condition
\begin{equation}
\frac{\delta}{\delta p_{{\rm st}}}\left[-\int p_{{\rm st}}\ln p_{{\rm st}}\:dxdv-\lambda\int p_{{\rm st}}E\:dxdv\right]=0,\label{eq:8}
\end{equation}
with Lagrange multiplier $\lambda=1/kT_{0}$, leading to $p_{{\rm st}}(x,v)=p_{{\rm eq}}\propto\exp(-E/kT_{0})$
with $E=mv^{2}/2+\phi(x)$. The canonical distribution can be identified
by a constrained maximisation of the informational entropy of the
system. It is significant to observe that the only constraint that
has to be taken into account involves the average of the energy, and
that this has its origin in the production of entropy in the reservoir
brought about by the dynamics of energy exchange between reservoir
and the system.  No constraints on other functions of energy should
be included in the procedure. Furthermore, the Lagrange multiplier
need not be deduced later on to make the result conform to the canonical
distribution; its value had been fixed when writing down the original
stochastic dynamical equations. Finally, it is easy to check using
Eq. (\ref{eq:ds3}) that $d\Delta s_{3}=0$ since $p_{{\rm eq}}(x,v)=p_{{\rm eq}}(x,-v)$,
and because $J_{v}^{{\rm ir,st}}=-\gamma vp_{{\rm eq}}-(\gamma kT_{0}/m)\partial p_{{\rm eq}}/\partial v=0$,
we can deduce from Eq. (\ref{eq:s2av}) that $d\langle\Delta s_{2}\rangle/dt=0$.
Moreover, consideration of Eq. (\ref{eq:a105}) leads to the stronger
conclusion $d\Delta s_{2}=0$, to be shown explicitly in the next
section.

\begin{figure}
\begin{centering}
\includegraphics[width=1\columnwidth]{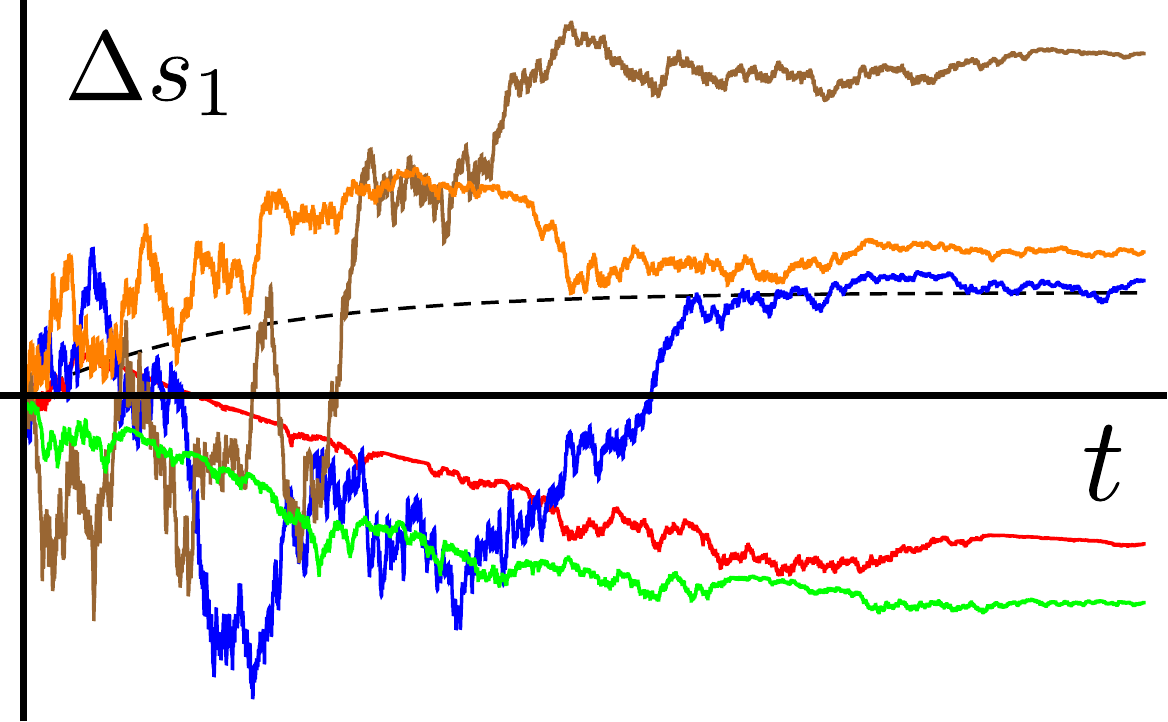}
\par\end{centering}

\caption{Examples of the time evolution of $\Delta s_{1}$ (continuous lines)
associated with phase space trajectories taken by a system as it relaxes
towards a stationary state for a particular stochastic process. The
dashed line is the average of $\Delta s_{1}$ over all possible trajectories,
satisfying $\langle\Delta s_{1}\rangle\ge0$. The stationary state
is approached as $t\to\infty$ and is characterised by the condition
$d\langle\Delta s_{1}\rangle/dt=0$ or equivalently the property $d\Delta s_{1}=0$
for all possible realisations, both of which are evident. \label{fig1}}
\end{figure}

An equivalent demonstration of the emergence of a canonical pdf from
the dynamics of stochastic entropy production is to note that $d\Delta s_{{\rm tot}}=d\Delta s_{1}=0$
for all possible incremental paths taken by the system when in the
equilibrium state, from which we conclude that $-d\ln p_{{\rm eq}}-dE/kT_{0}=0$
with the equilibrium pdf following by integration. The association
of $d\Delta s_{1}=0$ with the property $d\langle\Delta s_{1}\rangle/dt=0$
is illustrated in Figure \ref{fig1}.

Intuitively, the saturation of $S_{G}(t)-\overline{E}(t)/kT_{0}$
is equivalent to the maximisation of the uncertainty in the joint
microstate adopted by the system and reservoir, brought about by the
stochastic dynamics as $t\to\infty$. This is a reflection of the
progressive loss of knowledge of microstate, as time elapses, in models
where a system is coupled dynamically to a coarsely specified environment
instead of one where the microscopic detail is retained \cite{Ford-book2013}.

\section{Selection of a nonisothermal stationary state\label{sec:nonisothermal}}

\subsection{Underdamped dynamics\label{sub:Underdamped-dynamics}}

It is apparent that Eq. (\ref{eq:3}), together with the dynamics
of Eqs. (\ref{eq:1}) and (\ref{eq:1a}), can provide a framework
for the time evolution of the total entropy production for a system
exposed to an environment with a spatially dependent temperature.
We have up to now regarded $F$ as a conservative force, and this
could be generalised to include a nonconservative component that drives
a steady spatial flow. We investigate the latter in Appendix \ref{sec:Particle-in-a}
but in this section we shall restrict the discussion to a conservative
force field in order to make contact with a previous study of a particle
in a nonisothermal environment \cite{ZacHenry15a}. The system will
adopt a stationary nonequilibrium state where non-zero mean energy
flows take place between the reservoirs at each spatial position,
by way of the system, such that there will be a non-zero irreversible
current $J_{v}^{{\rm ir,st}}$ in the stationary state, bringing about
a steady rate of mean total entropy production. In the absence of
a nonconservative force we would expect the mean velocity in the stationary
state to vanish for all $x$.

We could identify the stationary pdf of such a system by solving the
appropriate Fokker-Planck equation, but there is an alternative entropy-based
approach. The dynamics generate a stationary pdf that embodies maximum
uncertainty at the microstate level consistent with the nonisothermal
constraints. The physical interpretation is that the component of
mean entropy production associated with relaxation evolves to become
as large as possible.

Our strategy is to establish an expression for $d\Delta s_{1}$ and
set it equal to zero for all dynamical scenarios, a condition equivalent
to $p=p_{{\rm st}}$ and hence through Eq. (\ref{S1av2}) to the reaching
of a ceiling in the mean value of $\Delta s_{1}$. Using the general
results in Appendix \ref{sec:Components-of-stochastic} for the dynamics
under consideration, we can write
\begin{eqnarray}
d\Delta s_{2} & = & \frac{kT_{r}\gamma}{m}\left(\frac{\partial\ln[1+\psi(x,-v)]}{\partial v}\right)^{2}dt\nonumber \\
 &  & -\frac{\partial\ln[1+\psi(x,-v)]}{\partial v}\left(\frac{2kT_{r}\gamma}{m}\right)^{1/2}dW,\label{eq:701}
\end{eqnarray}
where $\psi$ is a component of a convenient, but still general, specification
of the stationary pdf, namely $p_{{\rm st}}(x,v)=P_{{\rm st}}(x)[1+\psi(x,v)]f(x,v)$
where $f(x,v)=(m/2\pi kT_{r})^{1/2}\exp(-mv^{2}/2kT_{r})$ is a local
canonical distribution, and $P_{{\rm st}}=\int dv\,p_{{\rm st}}$.
Note that if $\psi=0$ we have a canonical distribution over velocity
and $d\Delta s_{2}$ vanishes as claimed in the previous section.

We also write
\begin{equation}
d\Delta s_{3}=-d\ln[1+\psi(x,v)]+d\ln[1+\psi(x,-v)],\label{eq:702}
\end{equation}
and using $d\Delta s_{1}=d\Delta s_{{\rm tot}}-d\Delta s_{2}-d\Delta s_{3}$
we have
\begin{eqnarray}
d\Delta s_{1} & = & -d\ln P_{{\rm st}}-d\ln f-\frac{1}{kT_{r}}d\left(\frac{mv^{2}}{2}\right)\nonumber \\
 &  & +\frac{F}{kT_{r}}dx-d\Delta s_{2}-d\ln[1+\psi(x,-v)],\label{eq:703}
\end{eqnarray}
and this reduces to
\begin{eqnarray}
d\Delta s_{1} & = & -d\ln P_{{\rm st}}+\left(1-\frac{mv^{2}}{kT_{r}}\right)\frac{T_{r}^{\prime}}{2T_{r}}dx+\frac{F}{kT_{r}}dx\nonumber \\
 &  & -\frac{kT_{r}\gamma}{m}\left(\frac{\partial\ln[1+\psi(x,-v)]}{\partial v}\right)^{2}dt\nonumber \\
 &  & -\frac{\partial\ln[1+\psi(x,-v)]}{\partial x}dx\nonumber \\
 &  & -\frac{\partial\ln[1+\psi(x,-v)]}{\partial v}\left(-\gamma v+\frac{F}{m}\right)dt\nonumber \\
 &  & -\frac{\partial^{2}\ln[1+\psi(x,-v)]}{\partial v^{2}}\frac{kT_{r}\gamma}{m}dt.\label{eq:704}
\end{eqnarray}
where $T_{r}^{\prime}=dT_{r}/dx$.

Setting $d\Delta s_{1}=0$ corresponds to a general condition for
the structure of $\psi$ and hence the pdf in the stationary state.
For illustration, however, we proceed with an assumption that $\psi$
is small and inversely proportional to $\gamma$, which is the commonly
used perturbative Chapman-Enskog representation \cite{Hirschfelder54,Cercignani-book2000},
a well-established approach to solving problems in kinetic theory
\cite{Lebowitz60} to first order in inverse friction coefficient.
This is not the same as making an assumption of overdamped dynamics,
which would involve a different specification of the underlying stochastic
dynamics. We shall consider such an approach in Section \ref{sub:Overdamped-dynamics}.
We identify the leading contributions to $d\Delta s_{1}$, namely
those of order $\gamma^{0}$ and write
\begin{eqnarray}
 &  & d\Delta s_{1}=-d\ln P_{{\rm st}}+\left(1-\frac{mv^{2}}{kT_{r}}\right)\frac{T_{r}^{\prime}}{2T_{r}}dx+\frac{F}{kT_{r}}dx\nonumber \\
 &  & +\frac{\partial\psi(x,-v)}{\partial v}\gamma dx-\frac{\partial^{2}\psi(x,-v)}{\partial v^{2}}\frac{kT_{r}\gamma}{m}dt+O(\gamma^{-1}).\quad\label{eq:705}
\end{eqnarray}
For the right hand side to vanish term by term we deduce that a polynomial
representation of $\psi(x,-v)$ can only contain linear and cubic
terms in $v$. We write $\psi(x,v)=av+cv{}^{3}$ and require contributions
to $d\Delta s_{1}$ proportional to $v^{2}dx$ to vanish by demanding
that
\begin{equation}
-\frac{mv^{2}}{kT_{r}}\frac{T_{r}^{\prime}}{2T_{r}}dx-3cv^{2}\gamma dx=0.\label{eq:706}
\end{equation}
so that $c=-mT_{r}^{\prime}/(6\gamma kT_{r}^{2})$.

We consider a situation where the particle is spatially confined by
the potential, in the sense that the pdf vanishes as $x\to\pm\infty$.
This implies the physical requirement that the mean velocity $\int_{-\infty}^{\infty}vp_{{\rm st}}(x,v)dv$
at a given $x$ is zero in the stationary state, equivalent to the
condition $\int_{-\infty}^{\infty}vf(x,v)\psi(x,v)dv=0$. This means
that $a+3ckT_{r}/m=0$ and hence to lowest order in $\gamma^{-1}$,
\begin{equation}
\psi(x,v)=\frac{T_{r}^{\prime}}{2\gamma T_{r}}\left(v-\frac{m}{3kT_{r}}v^{3}\right),\label{eq:707}
\end{equation}
from which we conclude that
\begin{equation}
\frac{\partial\psi(x,-v)}{\partial v}=-\frac{T_{r}^{\prime}}{2\gamma T_{r}}\left(1-\frac{m}{kT_{r}}v^{2}\right),\label{eq:708}
\end{equation}
and
\begin{equation}
\frac{\partial^{2}\psi(x,-v)}{\partial v^{2}}=\frac{mT_{r}^{\prime}}{\gamma kT_{r}^{2}}v,\label{eq:709}
\end{equation}
giving
\begin{equation}
d\Delta s_{1}=-d\ln P_{{\rm st}}+\frac{F}{kT_{r}}dx-\frac{T_{r}^{\prime}}{T_{r}}dx+O(\gamma^{-1}).\label{eq:710}
\end{equation}
We deduce that the spatial part of the stationary pdf, to lowest order
in $\gamma^{-1}$, is
\begin{equation}
P_{{\rm st}}\propto T_{r}^{-1}\exp\left(\int\frac{F}{kT_{r}}dx\right),\label{eq:711}
\end{equation}
and the selection procedure is complete, with the result
\begin{eqnarray}
p_{{\rm st}} & \propto & T_{r}^{-3/2}\exp\left(-\frac{mv^{2}}{2kT_{r}}+\int\frac{F}{kT_{r}}dx\right)\nonumber \\
 &  & \times\left[1+\frac{T_{r}^{\prime}}{2\gamma T_{r}}\left(v-\frac{m}{3kT_{r}}v^{3}\right)\right].\label{eq:712}
\end{eqnarray}
 The solution of the appropriate Fokker-Planck equation to first order
in $\gamma^{-1}$ should, of course, produce the same outcome \cite{ZacHenry15a},
but our purpose here is to demonstrate that it can emerge also from
considerations of entropy production.

We now examine the mean production of $\Delta s_{2}$, writing
\begin{equation}
\frac{J_{v}^{{\rm ir,st}}}{p_{{\rm st}}}=-\gamma v-\frac{kT_{r}\gamma}{mp_{{\rm st}}}\frac{\partial p_{{\rm st}}}{\partial v}=-\frac{kT_{r}\gamma}{m}\frac{\partial\ln(1+\psi)}{\partial v},\label{eq:50}
\end{equation}
and employing Eq. (\ref{eq:s2av}) we obtain
\begin{eqnarray}
\frac{d\langle\Delta s_{2}\rangle}{dt} & = & \int dxdv\;p\frac{kT_{r}\gamma}{m}\left(\frac{\partial\ln[1+\psi(x,-v)]}{\partial v}\right)^{2},\quad\label{eq:51}
\end{eqnarray}
which is consistent with a direct averaging of Eq. (\ref{eq:701}).
Inserting Eq. (\ref{eq:707}) we find that
\begin{eqnarray}
\frac{d\langle\Delta s_{2}\rangle}{dt} & = & \int dxdv\;p\frac{kT_{r}\gamma}{m}\left(\frac{T_{r}^{\prime}}{2\gamma T_{r}}\right)^{2}\left[1-\frac{m}{kT_{r}}v^{2}\right]^{2},\qquad\label{eq:51b-1}
\end{eqnarray}
to lowest order in $\gamma^{-1}$ and in the stationary state we therefore
have
\begin{equation}
\frac{d\langle\Delta s_{2}\rangle_{{\rm st}}}{dt}=\int dx\;P_{{\rm st}}\frac{kT_{r}^{\prime2}}{2m\gamma T_{r}}+O(\gamma^{-2}).\label{eq:51c-1}
\end{equation}
 Similarly, Eq. (\ref{eq:s3av}) gives
\begin{equation}
\frac{d\langle\Delta s_{3}\rangle}{dt}=-\int dxdv\;\frac{\partial p}{\partial t}\ln\left[\frac{1+\psi(x,v)}{1+\psi(x,-v)}\right],\label{eq:51a}
\end{equation}
which leads to
\begin{eqnarray}
\frac{d\langle\Delta s_{3}\rangle}{dt} & \approx & -\int dxdv\;\frac{\partial p}{\partial t}\left[\frac{T_{r}^{\prime}}{\gamma T_{r}}\left(v-\frac{m}{3kT_{r}}v^{3}\right)\right].\quad\label{eq:52-3}
\end{eqnarray}
This clearly vanishes in the stationary state, and for $p$ close
to stationarity in the velocity coordinate (in the sense that $p/P=f+O(\gamma^{-1})$,
where $P(x,t)=\int p\,dv$), the mean rate of production $d\langle\Delta s_{3}\rangle/dt$
is of order $\gamma^{-2}$.

Note that the stationary solution to the dynamics (\ref{eq:712})
satisfies a local equipartition condition $\int dv\,v^{2}p_{{\rm st}}=P_{{\rm st}}kT_{r}/m$
but that this relationship is valid only to first order in $\gamma^{-1}$,
in contrast to the approach of \cite{Polettini13} where such a condition
is taken to be a requirement to all orders. Under the dynamics assumed
here, the system is maintained away from exact local equipartition
through the flows of heat between the various local reservoirs.

\subsection{Overdamped dynamics\label{sub:Overdamped-dynamics}}

It is instructive to consider next the variational identification
of the stationary state under nonisothermal conditions within a framework
of \emph{overdamped} dynamics, and to contrast the outcome with the
analysis in Section \ref{sub:Underdamped-dynamics} for underdamped
dynamics. The revised dynamics will affect the form of each component
of entropy production, although it remains the case that the principal
relaxational component will increase on average until the stationary
state is reached.

It is well known \cite{vanKampen07,Gardiner09,Kupferman04} that Eqs.
(\ref{eq:1}) and (\ref{eq:1a}) reduce for large $\gamma$ to the
It$\bar{{\rm o}}$-rules SDE for the position coordinate:
\begin{equation}
dx=\frac{F(x)}{m\gamma}dt+\left(\frac{2kT_{r}(x)}{m\gamma}\right)^{1/2}dW,\label{eq:33-3}
\end{equation}
together with an associated Fokker-Planck equation for the positional
pdf $P^{{\rm od}}(x,t)$.  We investigate the entropy production
implied by these dynamics, focussing attention on $\Delta s_{1}^{{\rm od}}$
where the superscript `od' indicates that it is associated with the
overdamped dynamics.

For an It$\bar{{\rm o}}$-rules SDE $dx=A_{x}^{{\rm {\rm od}}}(x)dt+[2D_{x}^{{\rm od}}(x)]^{1/2}dW$,
Eq. (\ref{eq:a103}) in Appendix \ref{sec:Components-of-stochastic}
implies total entropy production given by
\begin{eqnarray}
 &  & d\Delta s_{{\rm tot}}^{{\rm od}}=-d\ln P^{{\rm od}}+\frac{A_{x}^{{\rm od}}}{D_{x}^{{\rm od}}}dx+\frac{dA_{x}^{{\rm od}}}{dx}dt-\frac{1}{D_{x}^{{\rm od}}}\frac{dD_{x}^{{\rm od}}}{dx}dx\nonumber \\
 &  & -\frac{A_{x}^{{\rm od}}}{D_{x}^{{\rm od}}}\frac{dD_{x}^{{\rm od}}}{dx}dt-\frac{d^{2}D_{x}^{{\rm od}}}{dx^{2}}dt+\frac{1}{D_{x}^{{\rm od}}}\left[\frac{dD_{x}^{{\rm od}}}{dx}\right]^{2}dt,\label{eq:34-1}
\end{eqnarray}
recognising that the rules of stochastic calculus now associated with
the variable $x$ differ from those that hold for the full dynamics
of Eqs. (\ref{eq:1}) and (\ref{eq:1a}) since $x$ evolves in Eq.
(\ref{eq:33-3}) under the direct influence of a stochastic term.
Inserting $A_{x}^{{\rm od}}=A_{x}^{{\rm {\rm od},ir}}=F/m\gamma$
and $D_{x}^{{\rm od}}=kT_{r}/m\gamma$ we find that
\begin{eqnarray}
d\Delta s_{{\rm tot}}^{{\rm od}} & = & -d\ln P^{{\rm od}}+\frac{F}{kT_{r}}dx+\frac{F^{\prime}}{m\gamma}dt-\frac{T_{r}^{\prime}}{T_{r}}dx\nonumber \\
 &  & -\frac{FT_{r}^{\prime}}{m\gamma T_{r}}dt-\frac{kT_{r}^{\prime\prime}}{m\gamma}dt+\frac{kT^{\prime2}}{m\gamma T}dt,\label{eq:35-1}
\end{eqnarray}
where $F^{\prime}=dF/dx$ and $T^{\prime\prime}=d^{2}T_{r}/dx^{2}$.
Since under It$\bar{{\rm o}}$ rules we have $d\left[\int FT_{r}^{-1}dx\right]=FT_{r}^{-1}dx+D_{x}^{{\rm od}}\left(F^{\prime}T_{r}^{-1}-FT_{r}^{-2}T_{r}^{\prime}\right)dt$
and $d\ln T_{r}=T_{r}^{-1}T_{r}^{\prime}dx+D_{x}^{{\rm od}}(d[T_{r}^{-1}T_{r}^{\prime}]/dx)dt$
this reduces to
\begin{equation}
d\Delta s_{{\rm tot}}^{{\rm od}}=-d\ln P^{{\rm od}}-d\ln T_{r}+\frac{1}{k}d\left[\int FT_{r}^{-1}dx\right].\label{eq:37-1}
\end{equation}
Since velocity coordinates are absent in the overdamped dynamics there
are no $d\Delta s_{3}^{{\rm od}}$ contributions and the confinement
of the particle to a potential such that there is no spatial current
in the stationary state would suggest that $d\Delta s_{2}^{{\rm od}}$
is zero as well. The observation that the overdamped dynamics miss
out the housekeeping entropy production in the stationary state is
the origin of an entropy anomaly \cite{Celani12,Ge14} between treatments
based on over- and underdamped dynamics, to be discussed shortly.

If the only non-zero contribution to $d\Delta s_{{\rm tot}}^{{\rm od}}$
is $d\Delta s_{1}^{{\rm od}}$, then the stationary state is specified
by $d\Delta s_{{\rm tot}}^{{\rm od}}=0$. From Eq. (\ref{eq:37-1})
we can therefore deduce the form of $P_{{\rm st}}^{{\rm od}}$ to
be
\begin{equation}
P_{{\rm st}}^{{\rm od}}(x)\propto T_{r}^{-1}\exp\int(F/kT_{r})dx,\label{eq:38-1}
\end{equation}
which may be confirmed as the stationary solution to the Fokker-Planck
equation
\begin{equation}
\frac{\partial P^{{\rm od}}}{\partial t}=-\frac{\partial}{\partial x}\left(\frac{FP^{{\rm od}}}{m\gamma}\right)+\frac{\partial^{2}}{\partial x^{2}}\left(\frac{kT_{r}P^{{\rm od}}}{m\gamma}\right)\label{eq:39-1}
\end{equation}
for the overdamped dynamics (\ref{eq:33-3}). We note that $P_{{\rm st}}^{{\rm od}}$
is consistent with the pdf in Eq. (\ref{eq:712}) obtained to $O(\gamma^{-1})$
using underdamped dynamics, when integrated over $v$.

This result allows us to check the assertion that $d\Delta s_{2}^{{\rm od}}$
vanishes. According to Eq. (\ref{eq:a105}) we write
\begin{eqnarray}
 &  & d\Delta s_{2}^{{\rm od}}=\frac{A_{x}^{{\rm od}}}{D_{x}^{{\rm od}}}dx+\frac{d\varphi^{{\rm od}}}{dx}dx\nonumber \\
 &  & -\frac{1}{D_{x}^{{\rm od}}}\frac{dD_{x}^{{\rm od}}}{dx}dx+\frac{1}{D_{x}^{{\rm od}}}\left(\frac{dD_{x}^{{\rm od}}}{dx}\right)^{2}dt+D_{x}^{{\rm od}}\left(\frac{d\varphi^{{\rm od}}}{dx}\right)^{2}dt\nonumber \\
 &  & -2\frac{d\varphi^{{\rm od}}}{dx}\frac{dD_{x}^{{\rm od}}}{dx}dt+A_{x}^{{\rm od}}\frac{d\varphi^{{\rm od}}}{dx}dt-\frac{A_{x}^{{\rm od}}}{D_{x}^{{\rm od}}}\frac{dD_{x}^{{\rm od}}}{dx}dt,\label{eq:39a-1}
\end{eqnarray}
where $\varphi^{{\rm od}}=-\ln P_{{\rm st}}^{{\rm od}}$. Clearly
we have $d\varphi^{{\rm od}}/dx=T_{r}^{\prime}/T_{r}-F/kT_{r}$ and
by inserting the appropriate $A_{x}^{{\rm od}}$ and $D_{x}^{{\rm od}}$
it follows that $d\Delta s_{2}^{{\rm od}}=0$.

It is possible to construct a MaxEnt principle for the selection of
$P_{{\rm st}}^{{\rm od}}$, in the form of a functional maximisation,
from Eq. (\ref{eq:37-1}) together with $d\Delta s_{{\rm tot}}^{{\rm od}}=d\Delta s_{1}^{{\rm od}}$
and $d\langle\Delta s_{1}^{{\rm od}}\rangle/dt\ge0$, namely
\begin{eqnarray}
 &  & \frac{\delta}{\delta P_{{\rm st}}^{{\rm od}}}\Biggl[-\int P_{{\rm st}}^{{\rm od}}\ln P_{{\rm st}}^{{\rm od}}\:dx-\int P_{{\rm st}}^{{\rm od}}\ln T_{r}\:dx\nonumber \\
 &  & +\int P_{{\rm st}}^{{\rm od}}(x)\left(\int^{x}dx^{\prime}\frac{F(x^{\prime})}{kT_{r}(x^{\prime})}\right)\:dx\Biggr]=0,\label{eq:40-2}
\end{eqnarray}
which is equivalent to the requirement that $S_{G}^{{\rm od}}-\overline{\ln T_{r}}+\overline{\int dx\,F(x)/kT_{r}(x)}$
should be maximised. The system informational entropy $S_{G}^{{\rm od}}$
is a functional of $P^{{\rm od}}$, in contrast to the more general
form $S_{G}$ given in terms of $p$. The second and third terms in
Eq. (\ref{eq:40-2}) are effective constraints on the maximisation
of system entropy for the selection of the stationary $P_{{\rm st}}^{{\rm od}}$
within a framework of overdamped dynamics, and they are unambiguous,
if more elaborate than the constraint that appears in the corresponding
isothermal case. They have their origin in contributions to the total
entropy production in the reservoirs. While demonstrating a connection
with the MaxEnt approach employed under isothermal conditions, such
a functional maximisation would perhaps not be the most natural approach
to use to select the pdf under nonisothermal conditions. Furthermore,
it has been derived only for overdamped dynamics. The better strategy
would be to focus on the condition for zero increment in $\Delta s_{1}$,
namely Eq. (\ref{eq:704}).

Let us now reflect on the differences in the thermodynamics that emerge
when we use overdamped rather than underdamped dynamics. As already
mentioned, a treatment using overdamped dynamics fails to capture
the housekeeping entropy production that is expected to take place
in a nonisothermal stationary state. For a system with an approximately
stationary velocity distribution, such that $p/P\approx f(1+\psi)=p_{{\rm st}}/P_{{\rm st}}$
we have seen that the mean contribution $d\langle\Delta s_{3}\rangle/dt$
is second order in $\gamma^{-1}$, and in the same circumstances we
can combine Eqs. (\ref{eq:5a}), (\ref{S1av2}) and (\ref{eq:51b-1})
to write
\begin{equation}
\!\frac{d\langle\Delta s_{{\rm tot}}\rangle}{dt}=-\!\!\int\!\!dx\frac{\partial P}{\partial t}\!\ln\!\left[\!\frac{P(x,t)}{P_{{\rm st}}(x)}\!\right]\!+\!\int\!\!dxP\frac{kT_{r}^{\prime2}}{2m\gamma T_{r}}+O(\gamma^{-2}).\label{eq:52-4-0}
\end{equation}
Replacing $P$ in the second term by $P_{{\rm st}}$, an approximation
valid if the system is close to stationarity, and inserting $\Delta s_{{\rm tot}}^{{\rm od}}=\Delta s_{1}^{{\rm od}}$,
we arrive at

\begin{equation}
\frac{d\langle\Delta s_{{\rm tot}}\rangle}{dt}=\frac{d\langle\Delta s_{{\rm tot}}^{{\rm od}}\rangle}{dt}+\int dx\;P_{{\rm st}}(x)\frac{kT_{r}^{\prime2}}{2m\gamma T_{r}}+O(\gamma^{-2}).\label{eq:52-4-1}
\end{equation}
This is the one dimensional version of a similar result in \cite{Celani12}
that highlighted an anomaly in mean entropy production between over-
and underdamped treatments of the dynamics of a system. It is clear
from our analysis that the additional term on the right hand side
of Eq. (\ref{eq:52-4-1}) is an approximate form of the mean housekeeping
entropy production that is captured by the underdamped treatment but
neglected when an overdamped dynamical model is adopted. A rational
basis for the difference is to be found in recognising that entropy
production is a consequence of the dynamics, and that modifications
in the construction of the equations of motion will introduce changes
in the form of the entropy production. The identification of the anomaly
as a mean housekeeping entropy production (to lowest order in $\gamma^{-1}$)
using the analysis presented here gives it a clear physical interpretation.

\section{Conclusions \label{sec:Conclusions}}

A framework of stochastic thermodynamics \cite{seifertprinciples}
provides a direct connection between entropy production and a stochastic
model of the trajectory-level dynamics of a system. This dynamical
connection has brought clarity to the concept of entropy production
in statistical physics, enabling it to be extended to individual realisations
of the evolution of a system and situations where fluctuations are
important. A description of the mean entropy production can emerge
from a treatment of the dynamics at the level of a Fokker-Planck equation,
namely the behaviour of the system probability density function, but
stochastic thermodynamics adds a crucial specification of the entropy
production in terms of the probabilities that certain trajectories
might be generated. We can separate the total production of entropy
into three components within this framework.

The MaxEnt procedure for selecting equilibrium probability density
functions has strong credentials as an exercise in logical inference,
but we argue that it is made more compelling by demonstrating that
it can arise naturally as a result of the underlying stochastic dynamics.
The central question of how to constrain the maximisation of the system
informational entropy is resolved by noting that the constraint terms
employed in the derivation of the canonical pdf can be related to
the mean entropy production in the environment. Constraints are therefore
associated with the dynamical couplings of the system to the environment,
and these often arise through the exchange of dynamically conserved
quantities. The underlying principle of MaxEnt is to maximise the
uncertainty in our perception of the microscopic state of the world.
This is underpinned by the monotonic increase and saturation of the
mean total entropy production at equilibrium according to the dynamics.

Such a framework can be extended to the selection of stationary pdfs
for nonequilibrium systems. It is a thermodynamic alternative to seeking
a time-independent solution to the appropriate Fokker-Planck equation
describing the dynamics. The average of the principal relaxational,
or nonadiabatic component of entropy production $\Delta s_{1}$ \cite{hatanosasa,adiabaticnonadiabatic0,SpinneyFord12a}
increases to a ceiling when the stationary state is reached, and increments
in this component thereafter vanish. By exploiting the latter property
the nonequilibrium pdf can be identified. We have implemented such
a strategy in a simple one dimensional system of trapped Brownian
motion in a thermal gradient using both under- and overdamped dynamics.
In doing so we recover the `anomaly' between the entropy production
obtained under the two treatments \cite{Celani12} and show that it
corresponds to housekeeping entropy production. We have also used
the strategy, in Appendix \ref{sec:Particle-in-a}, to recover the
known stationary pdf for a particle subjected to a constant nonconservative
force and isothermal conditions.

In summary, the principal insight presented in this paper is that
the stochastic dynamics of a system generate stationary statistics
in a fashion that maximises the mean of a certain component of entropy
production, $\Delta s_{1}$, and that this can map onto the procedure
of constrained maximisation of system informational entropy based
on logical inference. The procedure is equivalent to demanding that
the increment $d\Delta s_{1}$ brought about by the dynamics should
vanish for all possible trajectories. We have employed this approach
to select an equilibrium state of a system in an isothermal environment
and a nonequilibrium state under nonisothermal conditions. We also
consider an isothermal system subject to a nonconservative force.
The $d\Delta s_{1}=0$ condition is equivalent through Eq. (\ref{eq:ds1})
merely to the requirement that the pdf is stationary, but since we
can associate the evolution of $\Delta s_{1}$ with energy exchange
with an environment and associated system change during relaxation,
it has a physical interpretation \cite{hatanosasa}. $\Delta s_{1}$
is defined in Eq. (\ref{eq:5b}) in terms of probabilities of system
evolution at the level of trajectories, one of the foundational statements
of stochastic thermodynamics, from which the property $\langle\Delta s_{1}\rangle\ge0$
follows, and an algorithm for the maximisation of $\langle\Delta s_{1}\rangle$
is provided by the underlying system dynamics. We suggest that the
procedure provides a natural extension of canonical MaxEnt to nonequilibrium
situations, at least for systems governed by Markovian stochastic
dynamics.

\subsection*{Acknowledgements}

I am grateful for support from the Engineering and Physical Sciences
Research Council (EPSRC) Network Plus on Emergence and Physics far
from Equilibrium.

\appendix

\section{Components of stochastic entropy production\label{sec:Components-of-stochastic}}

We summarise the main results concerning the dynamics of components
of stochastic entropy production that are derived in more detail in
Spinney and Ford \cite{SpinneyFord12b}. For a system governed by
It$\bar{{\rm o}}$-rules Markovian stochastic differential equations
(SDEs)
\begin{equation}
d{\rm x}_{i}=A_{i}(\textbf{x},t)dt+B_{i}(\textbf{x},t)dW_{i},\label{eq:a100}
\end{equation}
where $\textbf{x}$ represents a set of dynamical variables $({\rm x_{1},{\rm x_{2},\cdots)}}$
such as $(x,v)$, we define
\begin{eqnarray}
A_{i}^{{\rm ir}}(\textbf{x},t) & = & \frac{1}{2}\left[A_{i}(\textbf{x},t)+\varepsilon_{i}A_{i}(\boldsymbol{\varepsilon}\textbf{x},t)\right]=\varepsilon_{i}A_{i}^{{\rm ir}}(\boldsymbol{\varepsilon}\textbf{x},t),\qquad\label{eq:a101}
\end{eqnarray}
\begin{equation}
A_{i}^{{\rm rev}}(\textbf{x},t)=\frac{1}{2}\left[A_{i}(\textbf{x},t)-\varepsilon_{i}A_{i}(\boldsymbol{\varepsilon}\textbf{x},t)\right]=-\varepsilon_{i}A_{i}^{{\rm rev}}(\boldsymbol{\varepsilon}\textbf{x},t),\quad\label{eq:a102}
\end{equation}
where $\varepsilon_{i}=1$ for variables ${\rm x}_{i}$ with even
parity under time reversal symmetry (for example position $x$) and
$\varepsilon_{i}=-1$ for variables with odd parity (for example velocity
$v$), and $\boldsymbol{\varepsilon}\textbf{x}$ represents $(\varepsilon_{1}{\rm x_{1},\varepsilon_{2}{\rm x_{2},\cdots)}}$.
Defining also $D_{i}(\textbf{x},t)=\frac{1}{2}B_{i}(\textbf{x},t)^{2}$,
it may be shown that the following It$\bar{{\rm o}}$-rules SDE for
the total entropy production emerges:

\begin{eqnarray}
 &  & d\Delta s_{{\rm tot}}=-d\ln p+\sum_{i}\biggl[\frac{A_{i}^{{\rm ir}}}{D_{i}}d{\rm x}_{i}-\frac{A_{i}^{{\rm rev}}A_{i}^{{\rm ir}}}{D_{i}}dt+\frac{\partial A_{i}^{{\rm ir}}}{\partial{\rm x}_{i}}dt\nonumber \\
 &  & -\frac{\partial A_{i}^{{\rm rev}}}{\partial{\rm x}_{i}}dt-\frac{1}{D_{i}}\frac{\partial D_{i}}{\partial{\rm x}_{i}}d{\rm x}_{i}+\frac{(A_{i}^{{\rm rev}}-A_{i}^{{\rm ir}})}{D_{i}}\frac{\partial D_{i}}{\partial{\rm x}_{i}}dt\nonumber \\
 &  & -\frac{\partial^{2}D_{i}}{\partial{\rm x}_{i}^{2}}dt+\frac{1}{D_{i}}\left(\frac{\partial D_{i}}{\partial{\rm x}_{i}}\right)^{2}dt\biggr],\quad\label{eq:a103}
\end{eqnarray}
where $p$ is the time dependent pdf of variables $\textbf{x}$. The
corresponding It$\bar{{\rm o}}$ SDE for the principal relaxational
entropy production is
\begin{equation}
d\Delta s_{1}=-d\ln p+\sum_{i}\biggl[-\frac{\partial\varphi}{\partial{\rm x}_{i}}d{\rm x}_{i}-D_{i}\frac{\partial^{2}\varphi}{\partial{\rm x}_{i}^{2}}dt\biggr],\label{eq:a104}
\end{equation}
where $\varphi=-\ln p_{{\rm st}}$ and $p_{{\rm st}}$ is the stationary
pdf. Also
\begin{eqnarray}
 &  & d\Delta s_{2}=\sum_{i}\biggl[-\frac{A_{i}^{{\rm ir}}A_{i}^{{\rm rev}}}{D_{i}}dt+\frac{A_{i}^{{\rm ir}}}{D_{i}}d{\rm x}_{i}+\varepsilon_{i}\varphi_{i}^{\prime}(\boldsymbol{\varepsilon}\textbf{x})d{\rm x}_{i}\nonumber \\
 &  & -\frac{1}{D_{i}}\frac{\partial D_{i}}{\partial{\rm x}_{i}}d{\rm x}_{i}+\frac{1}{D_{i}}\left(\frac{\partial D_{i}}{\partial{\rm x}_{i}}\right)^{2}dt+D_{i}(\varphi_{i}^{\prime}(\boldsymbol{\varepsilon}\textbf{x}))^{2}dt\nonumber \\
 &  & -2\varepsilon_{i}\varphi_{i}^{\prime}(\boldsymbol{\varepsilon}\textbf{x})\frac{\partial D_{i}}{\partial{\rm x}_{i}}dt+\varepsilon_{i}(A_{i}^{{\rm ir}}-A_{i}^{{\rm rev}})\varphi_{i}^{\prime}(\boldsymbol{\varepsilon}\textbf{x})dt\nonumber \\
 &  & -\frac{(A_{i}^{{\rm ir}}-A_{i}^{{\rm rev}})}{D_{i}}\frac{\partial D_{i}}{\partial{\rm x}_{i}}dt\biggr],\label{eq:a105}
\end{eqnarray}
specifies an increment in $\Delta s_{2}$, using notation $\varphi_{i}^{\prime}(\boldsymbol{\varepsilon}\textbf{x})=\varepsilon_{i}\partial\varphi(\boldsymbol{\varepsilon}\textbf{x})/\partial{\rm x}_{i}$
and
\begin{eqnarray}
d\Delta s_{3} & = & -d\ln p_{{\rm st}}(\textbf{x})+d\ln p_{{\rm st}}(\boldsymbol{\varepsilon}\textbf{x})\nonumber \\
 & = & \sum_{i}[\varphi_{i}^{\prime}(\textbf{x})\circ d{\rm x}_{i}-\varepsilon_{i}\varphi_{i}^{\prime}(\boldsymbol{\varepsilon}\textbf{x})\circ d{\rm x}_{i}],\label{eq:a106}
\end{eqnarray}
defines the third component. Stratonovich notation is used in the
second line for reasons of compactness, but a more elaborate It$\bar{{\rm o}}$-rules
version can be constructed.

For the dynamics specified by Eqs. (\ref{eq:1}) and (\ref{eq:1a})
we have $A_{x}^{{\rm ir}}=0$, $A_{x}^{{\rm rev}}=v$, $A_{v}^{{\rm ir}}=-\gamma v$,
$A_{v}^{{\rm rev}}=F/m$, $D_{x}=0$ and $D_{v}=kT_{r}\gamma/m$ and
using Eq. (\ref{eq:a103}) we recover Eq. (\ref{eq:3}).

\bigskip{}

\section{Particle in a nonconservative force field\label{sec:Particle-in-a}}

We consider how the stationary pdf of a particle evolving according
to Eqs. (\ref{eq:1}) and (\ref{eq:1a}) with a nonconservative constant
force field $F(x)=F_{0}$ and an isothermal environment $T_{r}(x)=T_{0}$
can be selected according to the condition that $\langle\Delta s_{1}\rangle$
should be maximised. As before, we regard this as synonymous with
$d\Delta s_{1}=0$.

By adapting Eq. (\ref{eq:704}), while demanding on physical grounds
that the pdf should be spatially independent, we replace $\ln[1+\psi(x,v)]$
by $h(-v)$ and write $p_{{\rm st}}(v)=P_{{\rm st}}\exp[h(-v)]f(v)$
with $f(v)=[m/(2\pi kT_{0})]^{1/2}\exp[-mv^{2}/(2kT_{0})]$. The function
$h$ is specified by
\begin{eqnarray}
d\Delta s_{1} & = & \frac{F_{0}}{kT_{0}}dx-\frac{kT_{0}\gamma}{m}\left(\frac{dh}{dv}\right)^{2}dt\nonumber \\
 &  & -\frac{dh}{dv}\left(-\gamma v+\frac{F_{0}}{m}\right)dt-\frac{d^{2}h}{dv^{2}}\frac{kT_{0}\gamma}{m}dt=0,\qquad\label{eq:704-2}
\end{eqnarray}
and the normalisation $\int dv\exp[h(-v)]f(v)=1$.  It is apparent
that a quadratic in $v$ is the highest finite polynomial form that
$h$ can take (the exponent $N$ in the leading term $v^{N}$ must
be even to preserve the normalisation but $N>2$ would produce a non-vanishing
term proportional to $v^{2(N-1)}$), and hence we write $h(v)=a_{0}+a_{1}v+a_{2}v^{2}$.
 The condition that terms in Eq. (\ref{eq:704-2}) proportional
to $v^{2}dt$ vanish is
\begin{equation}
-(kT_{0}\gamma/m)4a_{2}^{2}+2a_{2}\gamma=0,\label{eq:704-3}
\end{equation}
 with solutions $a_{2}=0$ or $m/(2kT_{r})$, but the normalisation
condition eliminates the second option. The condition for terms in
both $vdt$ and $dt$ to vanish is found to be  $a_{1}=-F_{0}/(\gamma kT_{0})$,
and  the normalised pdf is therefore specified by
\begin{equation}
h(v)=\frac{F_{0}^{2}}{2m\gamma^{2}kT_{0}}-\frac{F_{0}v}{\gamma kT_{0}},\label{eq:704-5}
\end{equation}
corresponding to
\begin{equation}
p_{{\rm st}}(v)\propto\exp\left(-\frac{m[v-F_{0}/(m\gamma)]^{2}}{2kT_{0}}\right),\label{eq:704-6}
\end{equation}
which is the known stationary state for such a system \cite{SpinneyFord12b}.
Adapting Eqs. (\ref{eq:701}) and (\ref{eq:702}) we identify the
increments in the remaining components of entropy production to be
\begin{eqnarray}
d\Delta s_{2} & = & \frac{kT_{0}\gamma}{m}\left(\frac{\partial h}{\partial v}\right)^{2}dt-\frac{\partial h}{\partial v}\left(\frac{2kT_{0}\gamma}{m}\right)^{1/2}dW\nonumber \\
 & = & \frac{F_{0}^{2}}{\gamma mkT_{0}}dt+\left(\frac{2F_{0}^{2}}{\gamma mkT_{0}}\right)^{1/2}dW,\label{eq:704-7}
\end{eqnarray}
and
\begin{equation}
d\Delta s_{3}=-dh(-v)+dh(v)=-\frac{2F_{0}}{\gamma mkT_{0}}dv,\label{eq:704-7-1}
\end{equation}
which do not vanish, even in the stationary state, unless $F_{0}=0$.
Distributions of the components of entropy production in a relaxation
process for this system were examined in \cite{SpinneyFord12b}.


%

\end{document}